\documentclass[a4paper,twoside]{article}
\usepackage{refs}
\newcommand{\affil}[1]{$^{\rm #1}$}

\usepackage[authoryear]{natbib}
\bibpunct{(}{)}{;}{a}{}{,}

\usepackage{graphicx}
\date{} %Please leave the date blank
\title{\large\bf\flushleft Outer Spiral Disks as Clues to Galaxy
Formation and Evolution}
\author{\parbox{\textwidth}{\flushleft
\vspace{-0.5cm}
{\it Marija Vlaji\'c\affil{A,B,C}}\\
\vspace{0.4cm}
{\small \affil{A}\,Astrophysics, Department of Physics, Keble Road,
University of Oxford, Oxford OX1 3RH, United Kingdom}\\ 
{\small\affil{B}\,Sydney Institute for Astronomy, School of Physics, 
The University of Sydney, NSW 2006, Australia}\\ 
{\small\affil{C}\,Email: vlajic@astro.ox.ac.uk}}}
\begin{document}
\twocolumn[
\begin{changemargin}{.8cm}{.5cm}
\begin{minipage}{.9\textwidth}
\vspace{-1cm}
\maketitle
%
%
%%%%%%%%%%%%%     ABSTRACT    %%%%%%%%%%%%%
%Abstract of no more than 200 words here.
\small{\bf Abstract:} Recent studies of outer spiral disks have given
rise to an abundance of new results.  We discuss the observational and
theoretical advances that have spurred the interest in disk outskirts,
as well as where we currently stand in terms of our understanding of
outer disk structure, ages and metallicities.

%%%%%%%%%%%%%     KEYWORDS    %%%%%%%%%%%%%
\medskip{\bf Keywords: galaxies: abundances --- galaxies: evolution
--- galaxies: spiral --- galaxies: stellar content --- galaxies:
structure}
% Please write all keywords in lower case. PASA uses the standard list
% of subject headings adopted by The Astrophysical Journal and
% available from
% http://www.journals.uchicago.edu/ApJ/keywords_text.html.  Keywords
% are separated by em-dashes, i.e. ---

%%%%%%%%DO NOT EDIT%%%%%%%%%%%%
\medskip
\medskip
\end{minipage}
\end{changemargin}
]
\small
%%%%%%%%EDIT FROM HERE%%%%%%%%%%%%

\defcitealias{roskar08a}{R08b}
\defcitealias{roskar08b}{R08a}
\defcitealias{sanchezblazquez09}{SB09}
\defcitealias{martinezserrano09}{MS09}

\section{Introduction}
%Please see the PASA Style Guide for help with correct layout for your
%manuscript.  Examples of tables and figures are given below.

The importance of studying faint outskirts of galaxies for our
understanding of the galaxy formation and evolution has become
increasingly apparent in recent years.  Due to their long dynamical
timescales, galactic outer regions have retained the fossil record
from the epoch of galaxy assembly in the form of spatial distribution,
kinematics, ages and metallicities of their stars
\citep{bullockjohnston05,freemanbh02}.  In addition, secular evolution
of spirals leaves the most conspicuous clues in the low density
regions in galactic outskirts
\citep{roskar08a,roskar08b,schoenrich08,sanchezblazquez09,martinezserrano09},
providing an opportunity for testing scenarios of galaxy evolution
using observations of outer disks of spirals.

In this paper we look into recent advances in observational studies
and numerical simulations of spiral disks, and the new insights they
have provided into the processes of disk galaxy formation and
evolution.

\section{Recent Advances in Outer Disk Studies}
\label{sec:advances}

The last decade has seen a growing interest in studies of faint outer
disks of spirals.  The progress made in the studies of disk outskirts
in the recent years has been brought together by a convergence of
observational advances and theoretical breakthroughs.  In this section
we discuss both of these aspects.

\subsection{Theoretical Advances}
\label{sec:theoretical}

\begin{figure}[!h]
\begin{center}
\hspace{-0.5cm}
\includegraphics[scale=0.3, angle=-90]{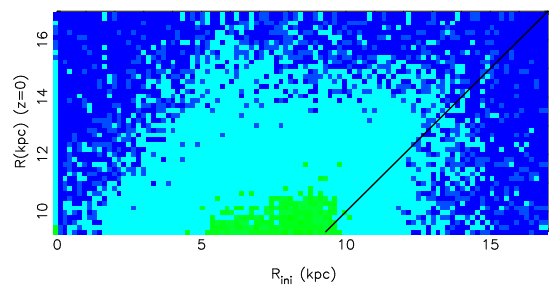}
\caption{Final stellar galactocentric radius versus formation radius.
From \citet{sanchezblazquez09}}\label{fig1}
\end{center}
\end{figure}

In 2002, Sellwood \& Binney demonstrated that a star just outside
(inside) the corotation radius can be caught in the spiral arm
resonance and transported inward (outward) while preserving the
circularity of its orbit \citep{sellwoodbinney02}.  They also showed
that transient spiral arms are at peak strength only for long enough
to produce a single crossing from one side of the corotation to the
other for each star.  As a consequence, a large number of stars are
scattered away from their birth radii (Figure~\ref{fig1}).  Recent
years have seen a renewed interest in studies of stellar radial
migration.  \citet[hereafter R08b]{roskar08a},
\citet[R08a]{roskar08b}, \citet[SB09]{sanchezblazquez09}, and
\citet[MS09]{martinezserrano09} explore high-resolution, N-body
simulations of disk formation and study the effects of stellar radial
mixing on the properties of the final disk galaxy.  A common feature
of these models are breaks in surface brightness at the radii of
$\sim3$ disk scale lengths, similar to what has been observed in a
large fraction of spirals \citep{pohlen06}.

Above simulations can be broadly divided into two classes -- those
that model an isolated disk galaxy \citepalias{roskar08a,roskar08b}
and those that study disk formation in the full cosmological context
\citepalias{sanchezblazquez09,martinezserrano09}.  Primary differences
between the two sets of models can be summarized as follows:

\noindent (i) While a \citetalias{roskar08a} galaxy experiences breaks
in both stellar surface density and surface brightness,
\citetalias{sanchezblazquez09} and \citetalias{martinezserrano09} find
no break in stellar surface density.  The latter suggests that the
break in the surface brightness profile is a consequence of a change
in stellar population properties across the break.
%points to a break in surface brightness that is a consequence of a
%change in stellar population properties across the break.

\noindent(ii) A drop in SFR, which in turn is due to a drop in the gas
surface density, is the cause of the break in surface brightness in
\citetalias{roskar08a}.  Similarly, the break in
\citetalias{sanchezblazquez09} is due to a decrease in the star
formation density per unit area, which itself originates from a
decline in volume density of gas at the break radius.  Interestingly,
this coincides with the radius at which the gas disk begins to warp
(although the causal relationship between the two effects has not been
proven).

\noindent (iii) In \citetalias{roskar08a} simulations, the outer disk
is populated solely by the stars migrated from the inner disk.
Migration accounts for a majority ($>60\%$) of stars found beyond the
break in \citetalias{sanchezblazquez09} and
\citetalias{martinezserrano09} galaxies, but star formation continues
at low level beyond the break.

\noindent (iv) All of the above works find a minimum in the stellar
age distribution at the break radius.  In \citetalias{roskar08a} this
is a consequence of radial migrations only, whereas
\citetalias{sanchezblazquez09} find that the minimum age at the break
radius is seeded by a break in star formation density and exists even
if migration in suppressed.

Stellar radial migrations work towards weakening the correlation
between metallicity and stellar age that is a clear prediction of
standard chemical evolution theory.  \citetalias{roskar08b} show that
radial mixing can indeed explain the apparent lack of the
age-metallicity relationship in the Galaxy; once the radial mixing has
been taken into account, predictions agree with the mostly flat
relationship observed in the Solar neighborhood.  Furthermore, radial
migrations have obvious consequences for the observed metallicity
distribution function and metallicity gradient, the issue we discuss in
more detail in Section~\ref{sec:metallicities}.

In addition to the effects on the observable properties of the Milky
Way, radial migrations could greatly influence our interpretation of
resolved stellar population observations in external spirals.
\citetalias{roskar08b} show that significantly different star
formation histories can be derived depending on whether migrations
have been taken into account or not.  The discrepancy is particularly
large in the outermost disk, strengthening the importance of outer
regions of spirals as testbeds for the models of galaxy evolution.

While the exact predictions of these models are still to be tested,
they could potentially dramatically change how we think about and
model galaxy evolution.

\subsection{Observational Advances}
\label{sec:observational}

\begin{figure}[!h]
\begin{center}
\hspace{-0.5cm}
\includegraphics[scale=0.4, angle=0]{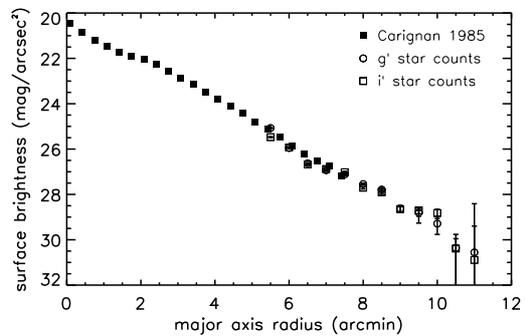}
\caption{Surface brightness profile of NGC~7793.  Full squares are
surface photometry data from \citet{carignan85}.  Open symbols are
effective surface brightness measurements derived from star counts.
From \citet{vlajic09b}}\label{fig2}
\end{center}
\end{figure}

Light distribution in spirals has been studied for many decades using
surface photometry.  These integrated-light studies have shown that
spiral disks generally follow an exponential surface brightness
profile \citep{devaucouleurs59,freeman70}, as well as that a fraction
of galaxies exhibits profiles characterized by a broken exponential
\citep{vanderkruit79,pohlen00,degrijs01,pohlen02,kregel02,kregel04,pohlen06}.
However, although a direct method, surface photometry suffers from
many technical difficulties at levels below $\mu_R\approx27$ mag
arcsec$^{-2}$.  These include difficulties of data flatfielding
\citep{pohlen02}, zodiacal light \citep{bernstein02}, diffuse Galactic
emission \citep{haikala95,reynolds92}, light scattered by the HI disk
\citep{blandhawthorn05} and effects of extended tails of the Point
Spread Function \citep{dejong08b}.  This limits the application of
surface photometry for studies of the extremely low surface brightness
regions.  It has been acknowledged that star counts offer a superior
method for tracing faint stellar populations in the outskirts of
galaxies.  By resolving individual stars and calculating the effective
surface brightness from star counts, it is possible to reach $3-5$ mag
arcsec$^{-2}$ deeper than with surface photometry
\citep{pritchet94,blandhawthorn05,irwin05,dejong07,ferguson07}.  The
example is shown in Figure~\ref{fig2}; in NGC~7793, star counts
profile significantly increases the known radial extent of the disk.

The most challenging aspect of studying outer disk light profiles is
the need to determine the background level that is then subtracted
from the raw profile in order to calculate the true star counts or
surface brightness profile of a galaxy.  In the case of surface
photometry this is manifested through difficulties in deriving
accurate estimates of sky brightness.  When resolved stellar
photometry is used instead to study outskirts of spirals, the task
translates into how to reliably evaluate the number of unresolved
faint background galaxies that are mistakenly included into stellar
catalogues.  The challenge is particularly difficult when information
on number counts as a function of color (and not only magnitude) is
required.  This problem can be -- and is being -- partially solved by
using space-based facilities with superior seeing, or ground-based
instruments with adaptive optics capabilities.  While difficulties in
surface photometry observations pertain to the nature of observations
themselves, star-galaxy separation which hinders resolved star
observations is limited by our knowledge of galaxy number counts and
ability to reliably distinguish between two classes of objects.

\section{Outer Disk Structure, Ages and Metallicities}
\label{sec:sam}

\subsection{Structure of Outer Disks}
\label{sec:structure}

A view of a simple exponential describing the light profile in spirals
\citep{freeman70} has evolved over the decades to include galaxies in
which an inner exponential is followed by an outer, steeper, one
\citep[a.k.a. broken, truncated, sub-exponential
profiles;][]{vanderkruit79,pohlen00,degrijs01,pohlen02,kregel02,kregel04,pohlen06}
as well as the galaxies with a shallower exponential following the
inner disk light profile \citep[a.k.a. upbending, anti-truncated,
super-exponential profiles;][]{erwin05}.  Both surface photometry and
star counts confirm the diversity in outer disk structure
\citep{blandhawthorn05,irwin05,ferguson07,barker07,dejong07}.
However, thirty years after its discovery, the origins of this
diversity are still not known.  Interactions and minor mergers have
been proposed to explain super-exponential profiles; this is supported
both by simulations \citep{penarrubia06,younger07} and by the finding
that these type of profiles are more common in high-density
environments \citep{pohlen07b}.  Scenarios proposed to explain
sub-exponential profiles can broadly be classified into those in which
the break is seeded during the galaxy formation processes \citep[and
marks the radius corresponding to the maximum angular momentum of the
proto-galaxy;][]{vanderkruit87}, and those in which a break in surface
brightness is a consequence of galaxy evolution \citep[having to do
with the threshold for star formation and/or secular evolution due to
spiral arms or
bars;][]{kennicutt89,schaye04,elmegreen06,debattista06,roskar08a,sanchezblazquez09,foyle08}.
\citet{vandenbosch01} combined the two scenarios to propose a picture
in which the distribution of cold gas is truncated to reflect the
maximum angular momentum of the collapsed cloud, while the break in
the stellar light profiles points to the presence of a star formation
threshold.  The ability of some spiral disks to retain the exponential
light profile out to very large distances
\citep[e.g. NGC~300,][]{blandhawthorn05} also remains a puzzle; in
fact, producing an extended purely exponential disk seems to be the
most challenging requirement for the simulations of disk formation and
evolution.

\subsection{Ages of Outer Disks}
\label{sec:ages}

Age behavior in spirals is very challenging to discern in galaxies
too distant for the full star formation history to be modeled.  This
is largely due to the age-metallicity degeneracy and the lack of
indicators that are primarily age-sensitive.  Rough overall age and
the dominant stellar population can however be relatively easily
determined from the color-magnitude diagram, and resolved stellar
photometry studies of nearby spirals find both young and old outer
disks.  For instance, \citet{davidge06,davidge07} find young and
intermediate-age stars at large galactocentric distances ($\sim7$ disk
scale lengths) in NGC~2403 and NGC~247, in agreement with an
inside-out growing stellar disk \citep{ryderdopita94,naabostriker06}.
On the contrary, M33, NGC~300, and NGC~7793 seem to harbor old outer
disks with red giant branch stars as a dominant stellar population
\citep{barker07,vlajic09,vlajic09b}.

As briefly mentioned in Section~\ref{sec:theoretical}, models of
galaxy formation that include the effects of secular evolution predict
a minimum stellar age at the break radius and a positive age gradient
in the outer disk.  \citetalias{sanchezblazquez09} interpret this
'U'-shaped age profile as being a consequence of the break in star
formation density, whereas in \citetalias{roskar08a} simulations the
particular shape of the age gradient arises exclusively due to stellar
migrations.  Two recent observational studies support these
theoretical predictions for the shape of the age gradient.

\citet{bakos08} derive surface brightness and color profiles from
surface photometry of a sample of 85 late-type galaxies from the SDSS
data.  They find that for galaxies which exhibit a break in their
surface brightness profiles, the $g'-r'$ color profile has a minimum
(i.e. bluest color) at the break radius and becomes redder in the
outer disk, mimicking the shape of the age gradient described above.
\citetalias{sanchezblazquez09} find the similar color gradient in
their simulated disk and confirm that the color profile indeed mirrors
the age (and not metallicity) gradient.

For galaxies that are sufficiently close, stellar color-magnitude
diagrams can be used to model detailed star formation history (SFH).
\citet{barker07} use this approach to derive SFHs of three fields in
the outer disk of M33 and find that beyond $9$ kpc mean stellar age is
an increasing function of radius.  Similarly, \citet{williams09}
reconstruct SFHs for four inner disk fields in M33.  Their fields span
the distances from $1$ to $6$ kpc and exhibit mean ages that decrease
from the center of the galaxy outward.  The two results point to an
overall picture in which the age gradient has a minimum at the break
radius \citep[$\sim9$ kpc in M33,][]{ferguson07}, similar to what has
been predicted by \citetalias{roskar08a},
\citetalias{sanchezblazquez09}, and \citetalias{martinezserrano09}.

\subsection{Metallicities of Outer Disks}
\label{sec:metallicities}

Traditional inside-out models for disk galaxy formation
\citep{ryderdopita94,naabostriker06} predict negative abundance
gradients in spirals.  Surface density, yield, and star formation all
decline with radius, resulting in metallicity distribution that is
more metal-rich in central parts and decreases progressively towards
the outer disk.  However, there is a growing body of evidence
suggesting that most spirals exhibit a flattening of their metallicity
gradient in the outermost disk.  Observationally, the strongest case
has been made for the Galaxy
\citep{andrievsky04,yong06,carraro07,pedicelli09}, M83
\citep{bresolin09}, and NGC~300 \citep{vlajic09}, although the effect
has been reported in a few additional spirals.  On the other hand, an
apparent lack of flattening is observed in NGC~7793 \citep{vlajic09b}
and possibly M33 \citep{barker07}.

Radial migrations offer a possible explanation for the shallower slope
of the outer disk abundance gradient.  \citetalias{roskar08b} find
that the slope of the stellar abundance gradient decreases with the
increasing age of stellar population, a result which is confirmed in
the \citetalias{sanchezblazquez09} work.  This, combined with the age
behavior described above results in the mean overall gradient which
flattens in the outer disk (at $\sim12$ kpc in
\citetalias{sanchezblazquez09}).

\section{Conclusions}

Thanks to a confluence of exciting new results from both observational
and theoretical work, outer spiral disks have in recent years become a
fast-growing research area.  While more deep stellar photometry is
necessary in order to test predictions of a growing body of numerical
simulations of disk formation and evolution, our current understanding
can be summarized as follows:

(i) Origin of the diversity in outer disk structure remains a puzzle.
While scenarios have been proposed to explain sub- and
super-exponential profiles, the full picture which self-consistently
explains existence of all three types of light profiles is lacking.

(ii) Radial migrations potentially play a significant role in the
evolution of disk galaxies; this could have profound consequences on
how we model galaxy evolution, in particular chemical evolution in
spirals.

(iii) Star counts have been shown to be a superior method for probing
faint outer disks in individual galaxies compared to traditional
surface photometry.

(iv) Age behavior in disks is very challenging to determine from
resolved stellar photometry.  However, a small number of studies seems
to indicate that in case of galaxies with sub-exponential profiles,
the minimum stellar age is observed at the break radius, in agreement
with simulations of radial mixing in spirals.

(v) Metallicity gradient which flattens in the outermost regions seems
to be a general feature of spiral disks.  However, some spirals
experience single-slope negative gradient with no flattening.

%%%Format tables as in the following example
%\begin{table}[h]
%\begin{center}
%\caption{Example Table}\label{tableexample}
%\begin{tabular}{lcc}
%\hline Column 1 & Column 2 & Column 3 \\
%\hline Table Content$^a$ \\
%\hline
%\end{tabular}
%\medskip\\
%$^a$Table footnotes go here.\\
%\end{center}
%\end{table}

\section*{Acknowledgments} %If needed

MV would like to thank Joss Bland-Hawthorn and Ken Freeman for their
collaboration and the organizers of the Galaxy Metabolism conference
for an interesting meeting.

%\begin{thebibliography}{}
%% References are listed as in the following example, for more examples,
%% please see the PASA Style Guide
%\bibitem[Smith, Jones, \& Brown(Year)Smith et al.]{example}Smith,
%A.~B., Jones, C.~D., Brown, E.~F. Year, Journal, Volume, Page
%\end{thebibliography}

%\bibliography{marija}
%\bibliographystyle{apj}

%\end{multicols}

\end{document}